\documentclass[prb,showpacs]{revtex4}
\usepackage{graphicx,psfrag,amssymb,amsmath}
\begin{document}
\title{The Fermi edge singularity in the SU(N) Wolff model}
\author{Bal\'azs D\'ora}
\email{dora@kapica.phy.bme.hu}
\affiliation{Department of Physics, Budapest University of Technology and 
Economics, H-1521 Budapest, Hungary}

\date{\today}

\begin{abstract}
The low temperature properties of the SU(N) Wolff impurity model are studied 
via Abelian bosonization. The path integral treatment of the problem allows for an exact
evaluation of low temperature properties of the model. The single particle
Green's function enhances due to the presence of local correlation. The basic correlation function such as
the charge or spin correlator are also influenced by the presence
of impurity, and show local Fermi liquid behaviour.
The X-ray absorption is affected by the presence of local Hubbard interaction. The exponent is decreased 
(increased) for repulsive (attractive) interactions.
\end{abstract}


\maketitle

\section{Introduction}

Local interactions have played an important role in explaining the behaviour of localized 
impurities, and in understanding the effect of electron correlation in lattice theories\cite{nersesyan}.
The most studied magnetic impurity models are the Anderson and Kondo model, including their various 
generalizations\cite{hewson}. Their local Fermi or non-Fermi liquid behaviour in the low temperature regime 
is 
inferred from Bethe ansatz method, but dynamical properties are still very difficult to be calculated.
Beyond the reliable renormalization group calculations\cite{muramatsu}, one is left with bosonization to 
gain insight into 
the low energy properties of such models.

One of the simplest description of a single impurity in a non-magnetic
metal is based on the Wolff model, where electron correlation is still present\cite{wolff,mattis,zhang}.
Also it is the basis of studying the effect of Coulomb interaction on resonant tunneling through a single
quantum level\cite{ng,oguri}. The model consists of conduction electrons interacting with each other 
\textit{solely} at 
the origin through the Hubbard interaction.
It is related to the Hubbard model in the same way as the single impurity Anderson model is
related to its periodic version.

Several extensions of this model are possible including extended interactions or 
by changing the properties of conduction electrons. In the present work we study a most natural 
generalization of the usual Wolff model: we consider fermions with SU(N) spin index or alternatively 
we take the ensemble of spin and orbital degrees of freedom into account by the N index
instead of the usual
two-valued spin index.
 The additional
degrees of freedom can be realized through orbital degeneracy, for
example, as in Mn oxides\cite{imada}. As a result, the additional degrees of freedom can be called flavour 
or color 
index.
In similar models, the inclusion of orbital or flavour index led to non-Fermi liquid behaviour, such as the 
two channel Kondo model\cite{emery,fabrizio}, or enriched our picture concerning the Mott-Hubbard 
transition as in the 
SU(N) Hubbard model\cite{assaraf,szirmai}.

The purpose of the present investigation is to see, how the interaction between N different species of 
electrons at a single site influence the low energy properties of the system. 
In the SU(N) Hubbard model with one particle per site, the Mott phase is reached at a finite value of the 
Hubbard interaction\cite{assaraf} for N$>$2, as opposed to the zero critical interaction for the SU(2) 
case\cite{nersesyan}. In the SU(2) Wolff model, the crossover to strong coupling region is predicted to be 
symmetric with respect to the sign of U\cite{mattis,zhang}. Here we shall see how this condition is 
modified 
for SU(N) symmetry.
 The use of Abelian 
bosonization allows for an exact evaluation of the basic correlation functions, which show the instability 
toward local magnetic moment formation (i.e. Kondo regime), as the interaction parameter is increased.
The critical interaction for repulsive U is independent of the value of N, but for attractive interaction 
its value is suppressed by 1/(N-1).
The low energy behaviour of the system is characterized by a local Fermi liquid, obeying the well-known 
Fermi liquid relations (such as the Korringa relations) of the Anderson impurity model. 

The X-ray response of the SU(N) Wolff model is evaluated with the use of boundary condition changing 
operators\cite{affleck} and is qualitatively similar to a system with correlated electrons over the 
whole lattice (namely the Hubbard model)\cite{affleck}. The exponent characterizing the scaling dimension 
of the 
deep level electrons decreases with increasing Hubbard interaction.
 
\section{Formalism, model Hamiltonian}

The Hamiltonian describing N different species of electrons interacting
only at the origin is given by:
\begin{gather}
 H=\sum_{m=1}^N\left[-iv\int\limits_{-\infty}^\infty dx
\Psi^+_m(x)\partial_x\Psi_m(x)+E:\Psi^+_m(0)\Psi_m(0):+\frac U2
\sum_{n=1,n\neq m}^N:\Psi^+_m(0)\Psi_m(0)::\Psi^+_n(0)\Psi_n(0):\right],
 \label{hamilton}
\end{gather}
and only the radial motion of the particles is accounted for by chiral (right moving) fermion 
fields\cite{affleck2}.
The model can be bosonized via\cite{nersesyan,delft}
\begin{equation}
\Psi_m(x)=\frac{1}{\sqrt{2\pi\alpha}}e^{i\sqrt{4\pi}\Phi_m(x)},
\end{equation}
and the bosonized Hamiltonian takes the form as
\begin{equation}
H=v\int\limits_{-\infty}^\infty dx \sum_m (\partial_x\Phi_m(x))^2+\frac E{\sqrt\pi}\sum_m
\partial_x\Phi_m(0)+\frac{U}{2\pi}\sum_{n,m=1,n\neq m}^N\partial_x\Phi_m(0)\partial_x\Phi_n(0).
\end{equation}
By introducing charge and spin fields as\cite{assaraf}
\begin{gather}
\Phi_c(x)=\frac{1}{\sqrt{N}}\sum_{m=1}^N\Phi_m(x),\\
\Phi_{n,s}(x)=\frac{1}{\sqrt{n(n+1)}}\left(\sum_{m=1}^n\Phi_m(x)-n\Phi_{n+1}(x)\right),
\end{gather}
$n=1$\dots$N-1$, the Hamiltonian separates into different sectors: the spin sector is described by $N-1$ identical
decoupled bosonic modes, and the charge sector transforms into a similar
massless bosonic mode as
\begin{gather}
H_{s}=\sum_{m=1}^{N-1}\left[v\int\limits_{-\infty}^\infty dx 
(\partial_x\Phi_{m,s}(x))^2-\frac{U}{2\pi}(\partial_x\Phi_{m,s}(0))^2\right],\\
H_c=v\int\limits_{-\infty}^\infty dx (\partial_x\Phi_c(x))^2+E\sqrt\frac N\pi 
\partial_x\Phi_c(0)+\frac{(N-1)U}{2\pi}(\partial_x\Phi_c(0))^2.
\end{gather}
Since the Hamiltonians describe scattering of massless bosons on a point-like
``impurity'', the bosonic correlation functions can readily be
calculated. 
Hence in the followings it suffices to study the single Hamiltonian
\begin{equation}
H=v\int\limits_{-\infty}^\infty dx (\partial_x\Phi(x))^2+E
\partial_x\Phi(0)+U(\partial_x\Phi(0))^2
\label{model}
\end{equation}
describing both sectors with the appropriate parameters.
When evaluating the generating functional in terms of path integral, one can benefit 
from the fact that the 
Hamiltonian depends only on $\partial_x\Phi(x)$, and not separately on the dual fields $\phi(x)$ and 
$\theta(x)$. As a result, the path integral can be formulated after identifying the "space 
coordinate" with 
$\Phi(x)$ and the "momentum" conjugate to it\cite{jackiw} with $-\partial_x\Phi(x)$ from the commutation 
relation of the $\Phi$ field with itself
($[\Phi(x),\Phi(y)]=i\textmd{sgn}(x-y)/4$). The obtained action is for
chiral bosons, similar to the one investigated during the study of edge
states\cite{nersesyan}.
Then the generating functional is obtained as 
\begin{equation}
Z\sim\int D\Phi\exp(S(\Phi))
\end{equation}
with
\begin{gather}
S=\int\limits_0^\beta d\tau'\left[ 
-i\int\limits_{-\infty}^\infty \partial_{\tau'}\Phi(x,\tau')\partial_x\Phi(x,\tau')
-v\int\limits_{-\infty}^\infty dx 
(\partial_x\Phi(x,\tau'))^2-E\partial_x\Phi(0,\tau')-z(\tau')^2-2z(\tau')\sqrt{-U}\partial_x\Phi(0,\tau')\right]+\nonumber\\
+i\mu(\Phi(0,\tau)-\Phi(0,0)),
\end{gather}
where the interaction term was decoupled by a time dependent Hubbard Stratonovich field $z(\tau)$. For 
$\mu=0$, it gives the partition function, for $\mu=\sqrt{4\pi/N}$ it equals to the local Green's function at 
the impurity site ($G(\mu)=Z(\mu)/Z(0)$) coming from either the charge or one of the spin sectors.
After Fourier transforming the fields as
\begin{gather}
\Phi(x,\tau)=\sum_{m,q}\exp(-i\omega_m\tau-iqx)\Phi(m,q),\\
z(\tau)=\sum_{m}\exp(-i\omega_m\tau)z(m),
\end{gather}
$\omega_m=2m\pi T$ is the bosonic Matsubara frequency, 
the action is transformed to
\begin{gather}
S=\sum_{p}(-iq\omega_m-vq^2)\Phi(-p)\Phi(p)+iE\sum_q q\Phi(0,q)-\sum_mz(-m)z(m)+2i\sqrt{-U}
\sum_p z(-m)q\Phi(p)+\nonumber \\
+\sum_p i\mu(\exp(-i\omega_m\tau)-1)\Phi(p),
\end{gather}
where $p=\{m,q\}$. The $\Phi$ field can be integrated out, and the effective action reads
as
\begin{equation}
S_{eff}=-\sum_mz(-m)z(m)+\frac 12\sum_p\eta(-p)G\eta(p),
\end{equation}
where 
\begin{gather}
G^{-1}=vq^2+iq\omega_m,\\
\eta(p)=iEq\delta_{m,0}+2i\sqrt{-U}z(-n)q+i\mu(\exp(-i\omega_m\tau)-1).
\end{gather}
Integrating over the $z$ field gives for the free local Green's function
(where one has to substitute $\mu=\sqrt{4\pi/N}$) in the large $\tau$ limit
\begin{equation}
G(\tau,0)\sim \exp\left(-\frac{\mu^2 T}{2}\sum_{p}\frac{1-\cos(\omega_m\tau)}{vq^2+iq\omega_m}\right)\sim
\left(\frac{T}{\sin(\pi T\tau)}\right)^{\mu^2/4\pi}.
\label{green1}
\end{equation}
As to the $U$ dependent part, it is obtained as
\begin{equation}
G_{corr}(\tau,0)\sim\exp\left(-U\mu^2T\sum_m
|\gamma(\omega_m)|^2\frac{1-\cos(\omega_m\tau)}{1+U\Lambda(\omega_m)}\right)\sim
\left(\frac{U_0}{U_0+2\pi U}\right)^{\mu^2/2\pi}
\label{green2}
\end{equation}
in the large $\tau$ limit with 
\begin{gather}
\gamma(\omega_m)=\sum_q\frac{2}{i\omega+vq},\\
\Lambda(\omega_m)=\sum_q\frac{2q}{i\omega+vq},
\end{gather}
$U_0=\pi v/2n_0$, $n_0$ is the average density per spin channel in the
homogeneous case.
For $\mu=0$, the change of the free energy of Eq. (\ref{model}) in the presence of $U$ is evaluated as
\begin{equation}
F=T\sum_m\ln(1+U\Lambda(\omega_m))-\frac 14\frac{E^2\Lambda(0)}{1+U\Lambda(0)},
\end{equation}
which gives for the total specific heat
\begin{equation}
C(T)=\frac{\pi T}{6 v} \frac{U_0}{U_0+2\pi U}.
\label{fajho0}
\end{equation}
The presence of local Hubbard $U$ enhances the Sommerfeld coefficient for
attractive interaction, and decreases it for repulsive one.

Another way to derive the same results would have been to follow the diagrammatic approach.
Due to the presence of a single scattering center, the determinations and solution
of the corresponding Dyson equation would have led to the same exact expressions for the 
Green's function and free energy. 

\section{Correlation functions of the SU(N) Wolff model}

\begin{figure}[h!]
\psfrag{x}[t][b][1][0]{$U/U_0$}
\psfrag{y}[b][t][1][0]{$C_W(T)/C(T)$}
\centering{\includegraphics[width=7cm,height=7cm]{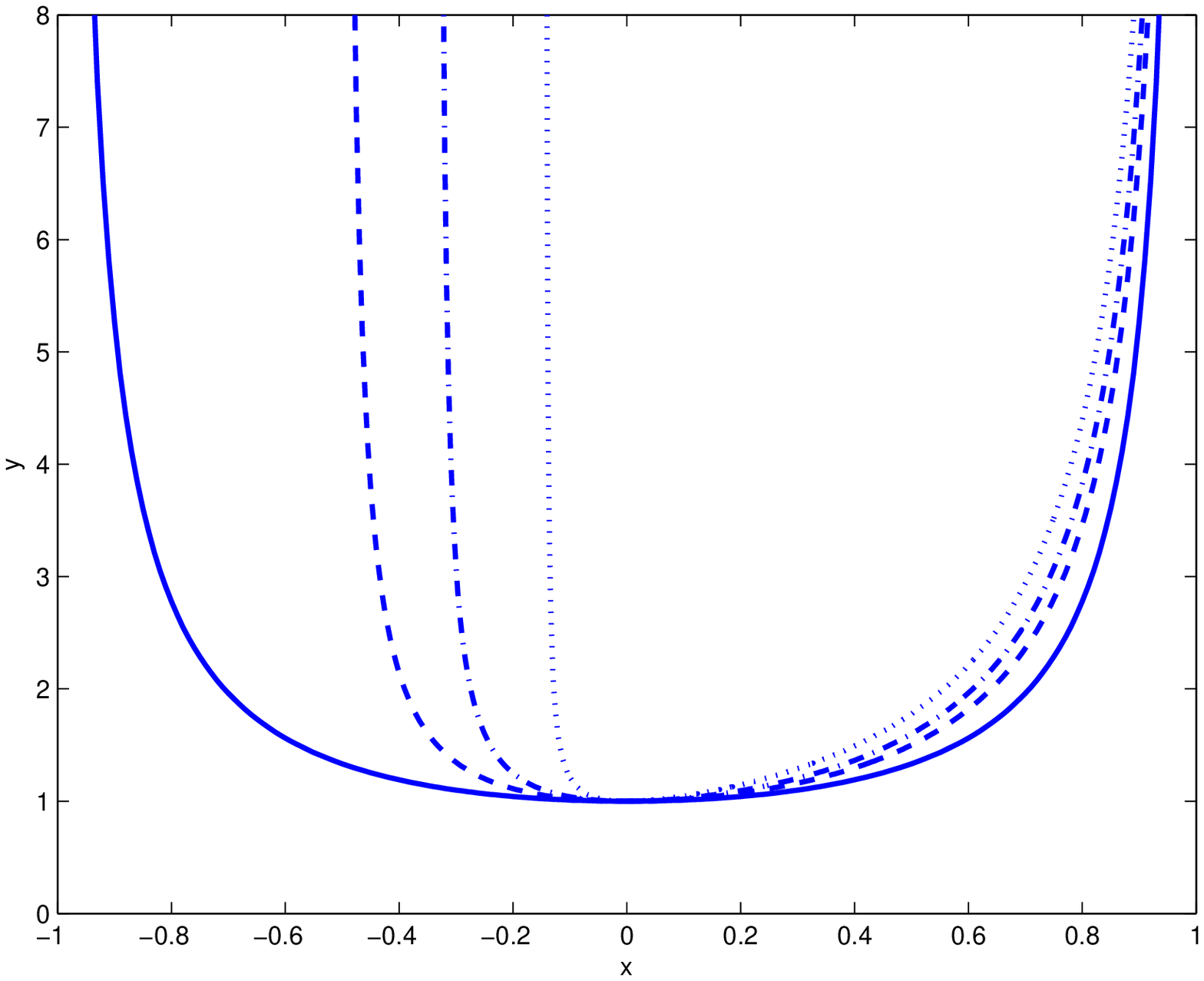}}
\hspace*{1cm}
\psfrag{x}[t][b][1][0]{$U/U_0$}
\psfrag{y}[b][t][1][0]{Density of states at the Fermi energy}
\centering{\includegraphics[width=7cm,height=7cm]{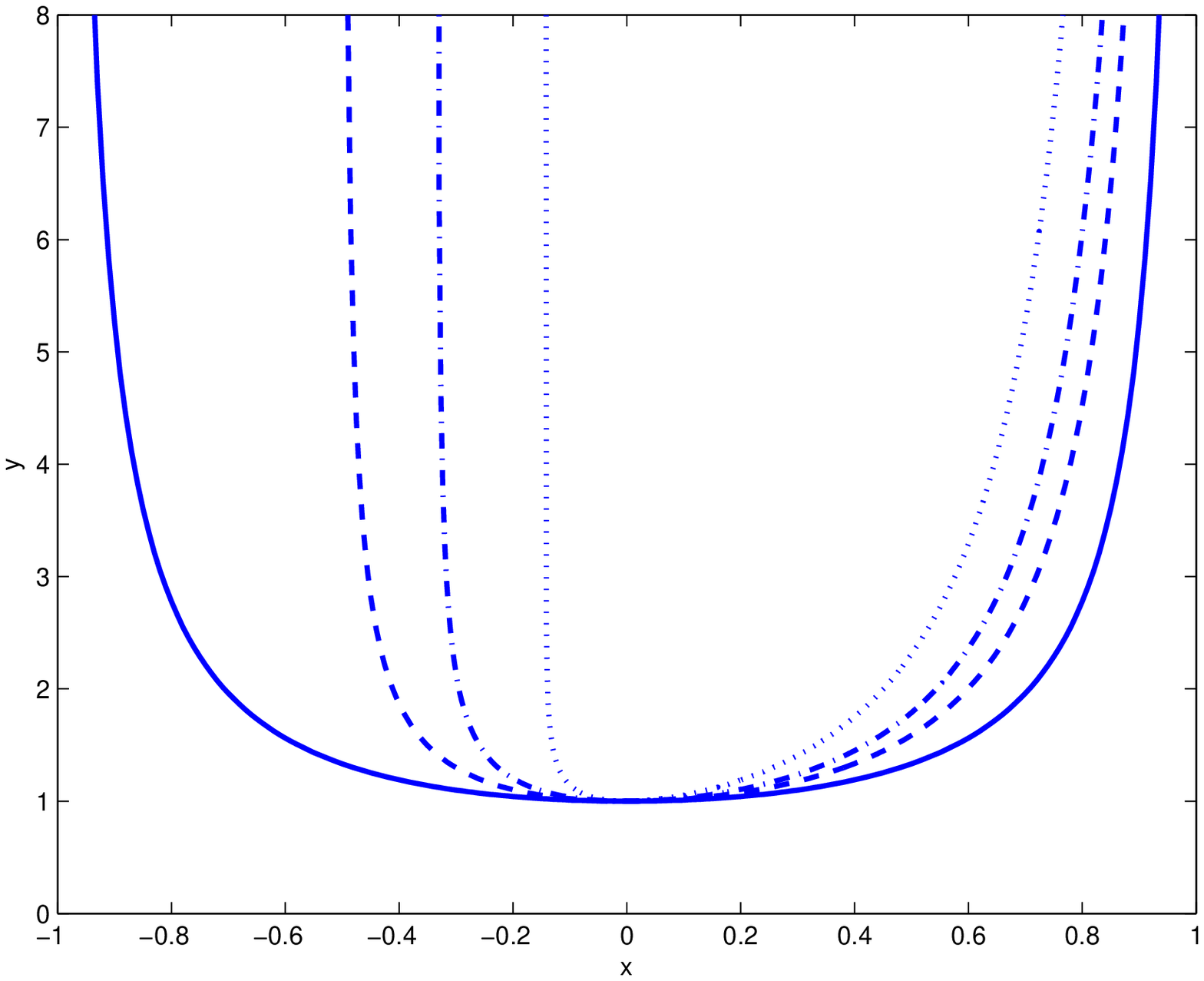}}
\caption{The Sommerfeld coefficient (left panel) and the density of states at the Fermi energy (right panel) 
are plotted as a function of local Hubbard
$U$ for $N=2$, $3$, $4$ and $8$ from edge to center. As is seen, the critical value of the interaction depends on its sign, namely $U_c=U_0$
for repulsion and $-U_c=U_0/(N-1)$ for attraction.\label{dos0}}
\end{figure}

By inserting the appropriate parameters into these
results, we are able to determine the low temperature-low energy properties
of Eq. (\ref{hamilton}).
First of all, the specific heat including the impurity contribution reads as
\begin{equation}
C_{W}(T)=\frac{\pi T}{6 v}\left(\frac{U_0}{U_0+(N-1)U}+\frac{(N-1)U_0}{U_0-U}\right),
\end{equation}
as opposed to its normal state value $C(T)=N\pi T/6v$.
In Fig. \ref{dos0}, we show the Sommerfeld coefficient, $C_W(T)/C(T)$ as a function of $U$.  The main quantity of interest is the local electron propagator. Putting the above results together,
one finds for the single particle electron Green's function at the impurity site in the 
$0\ll\tau\ll 1/T$ limit
\begin{equation}
G_m(\tau)=-\langle\Psi_m(\tau,0)\Psi_m^+(0,0)\rangle=-\frac{T}{\sin(\pi
T\tau)}\frac{ U_0^{2}}{(U_0+(N-1)U)^{2/N}(U_0-U)^{2-2/N}},
\label{green}
\end{equation}
where $U_0=\pi v/2n_0$ is the critical Hubbard interaction in the SU(2) Wolff impurity model, 
$n_0$ is the average density in the homogeneous system. For general
N$>$2, however, 
the critical value of $U$ is suppressed by
1/N as seen from the denominator of Eq. \ref{green}. When $U$
exceeds  this value, we expect local magnetic 
moment formation, and bosonization breaks down\cite{mattis,zhang}.
The one-body potential, $E$ does not appear in Eq. (\ref{green}) to leading order in $\tau$,
only renormalizes the $\tau^{-2}$ correction. 
 The $\tau\rightarrow\infty$ limit determines the low energy behaviour
of the local density of states. It remains constant, i.e. Fermi liquid like, but
enhances due to the presence of local Hubbard $U$, as can be seen in Fig. \ref{dos0}.

In the SU(2) case, the Sommerfeld coefficient and the local residual density of states have the same form, because the
Hamiltonian splits into two sectors somehow corresponding to the original two species of electrons. 
For N$>$2, however, we have N different species, but only have two distinct sectors with different weights: N-1 for the spin 
sector as opposed to the single charge sector, resulting in different behaviour in the specific heat and Green's function.
  
\begin{figure}[h!]
\psfrag{x}[t][b][1][0]{$\omega/W$}
\psfrag{y}[b][t][1][0]{Im$\chi_{charge}(\omega)/\chi_{charge}(0)$}
\centering{\includegraphics[width=7cm,height=7cm]{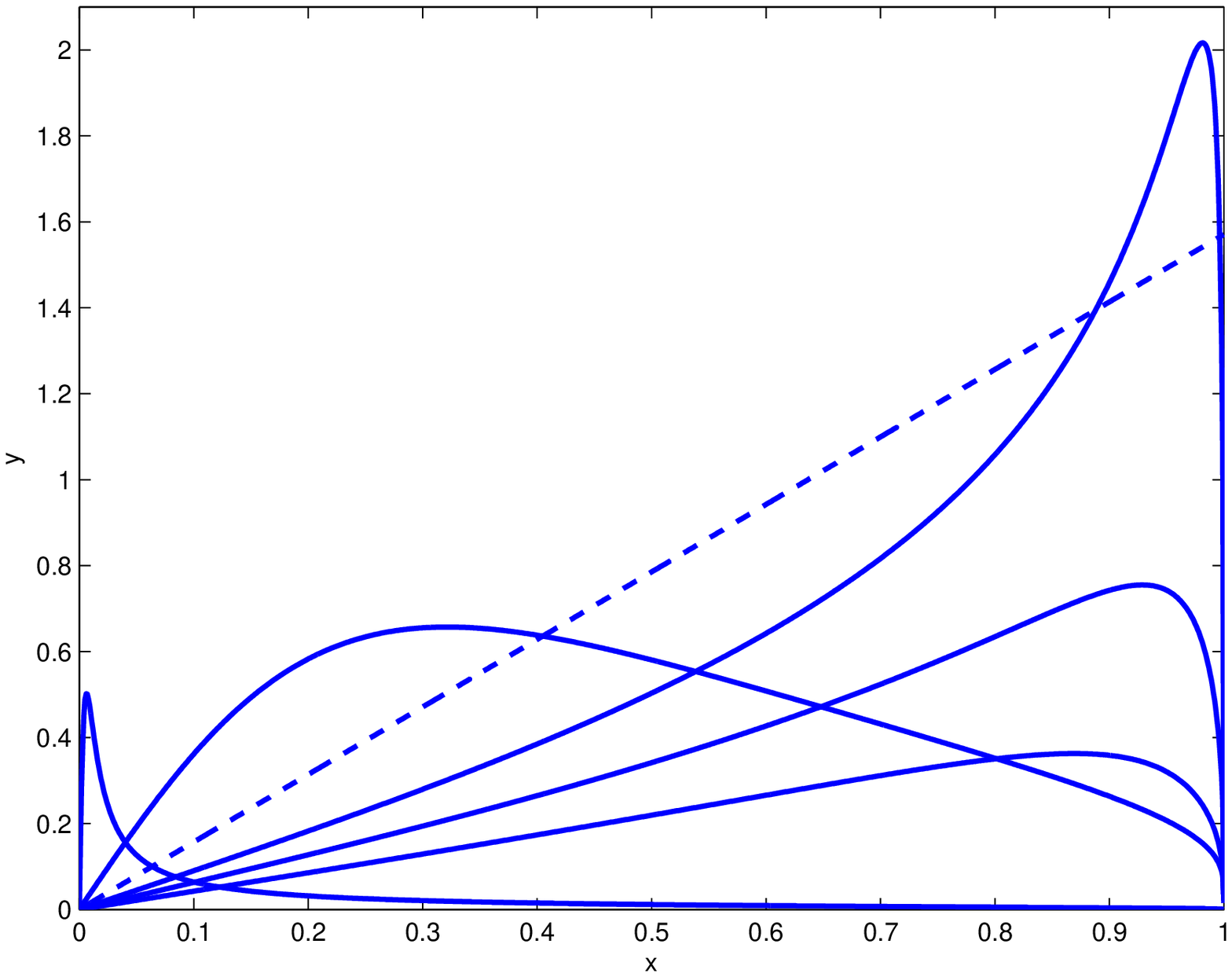}}
\hspace*{1cm}
\psfrag{x}[t][b][1][0]{$\omega/W$}
\psfrag{y}[b][t][1][0]{Im$\chi_{spin}(\omega)/\chi_{spin}(0)$}
\centering{\includegraphics[width=7cm,height=7cm]{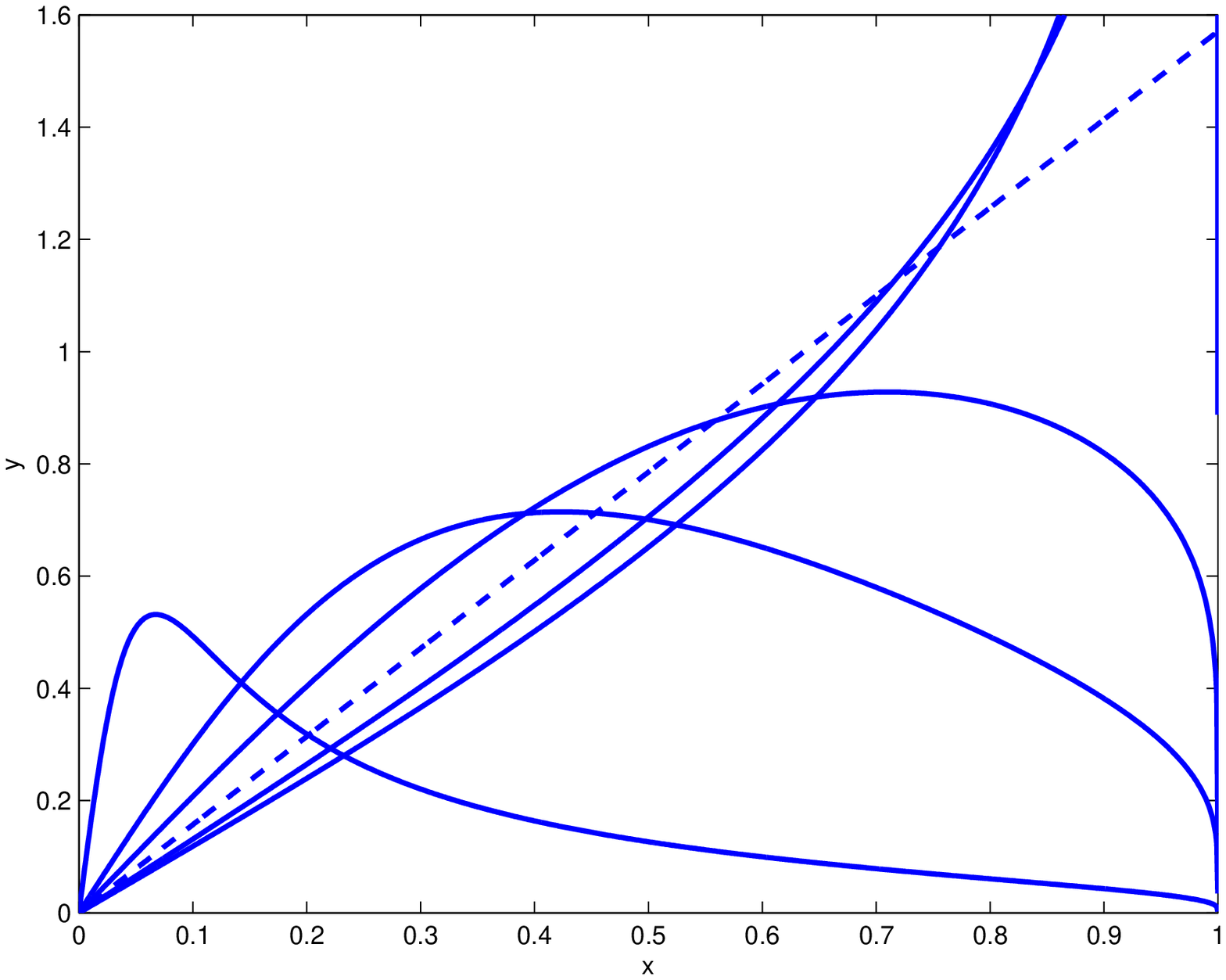}}
\caption{The imaginary part of the density-density response function (left panel) and the imaginary part of the spin
susceptibility are shown as a function of
frequency for N=4, $U=0.9U_0$, $0.5U_0$, $0.25U_0$, $0$ (dashed line), $-0.2U_0$ and $-0.33U_0$ with 
increasing (decreasing) initial slope at $\omega=0$ for the charge (spin) response. 
$W$ is the bandwidth assuming a sharp momentum space cut-off.\label{denss}}
\end{figure}

The other, $\tau\rightarrow 0^+$ 
limit determines the particle density as
\begin{equation}
G_m(\tau\rightarrow 0^+)=n_0-\frac{\sqrt N E}{U_0+(N-1)U}.
\end{equation}
The computation of the local density-density correlation function leads to 
\begin{equation}
\chi_{charge}(\omega)=\frac{N\chi_0(\omega)}{1+(N-1)U\chi_0(\omega)},
\end{equation}
where $\chi_0(\omega)$ is the local density correlation function in the
homogeneous case, $\chi_0(0)=1/U_0$. Due to Ward identity\cite{schlottmann}, the exact susceptibility is
identical to the RPA result in the SU(N) case as well\cite{zhang}.
The above result can further be verified by the fluctuation-dissipation theorem. Namely, the charge 
response function is related to the charge correlator by $\langle n^2\rangle-\langle n 
\rangle^2=T\chi_{charge}$. Here $n$ is the charge density, and its momenta can be obtained 
by
deriving the partition function with respect to $E$. 
At low frequency, where bosonization is expected to be valid\cite{nersesyan}, it is well approximated by
\begin{equation}
\chi_{charge}(\omega)\approx\frac{N}{U_0+(N-1)U}+i\frac{\omega}{2\pi v^2}\frac{NU^2_0}{(U_0+(N-1)U)^2}.
\label{chkis}
\end{equation}
Its frequency dependence is shown in Fig. \ref{denss} for N=4 and for various values of $U$. As is seen, 
the low frequency part is influenced more significantly for attractive interactions. As $U$ decreases below 
zero, its imaginary part becomes a Dirac delta function at the origin, signaling the breakdown of bosonization.
Similarly, the susceptibility in the spin sector can be evaluated as
\begin{equation}
\chi_{spin}(\omega)=\frac{(N-1)\chi_0(\omega)}{1-U\chi_0(\omega)},
\end{equation}
which again reduces to the RPA results. This is N-1 times the spin susceptibility of the
SU(2) Wolff model, stemming from the N-1 massless bosonic modes in the spin
sector.
The low energy spin response is evaluated as
\begin{equation}
\chi_{spin}(\omega)\approx\frac{N-1}{U_0-U}+i\frac{\omega}{2\pi v^2}\frac{(N-1)U^2_0}{(U_0-U)^2}.
\label{spkis}
\end{equation}
Compared to Eq. (\ref{chkis}), for repulsive interaction the spin response
increases faster in energy than the charge one due to the denominator
Eq. (\ref{spkis}), as can be checked directly in Fig. \ref{denss}. Similar results were found in the SU(2) case\cite{zhang}. 
Exactly at the critical $U$ deduced
from the Green's function, the charge or spin susceptibility diverges,
signaling a possible transition or crossover to  strong coupling case.

\section{X-ray absorption}

The basic Hamiltonian describing our previously studied system and deep level
electrons ($d$) is given by 
\begin{equation}
H_{deep}=H+E_0\sum_{n=1}^Nd_n^+d_n+V\sum_{m,n=1}^N\Psi^+_m(0)\Psi_m(0)d_n^+d_n,
\end{equation}
and the coupling of the metal to the X-ray field, describing the transfer
of a deep level electron to the conduction band is given by
\begin{equation}
H_X=W\sum_m\Psi_m^+(0)d_me^{-i\omega t}+h.c.
\end{equation}
The quantity of interest is the X-ray scattering rate, which is written in
the form of the Fermi golden rule as
\begin{gather}
I(\omega)=2|W|^2\textmd{Im}S(\omega),\\
S(t)=\langle d_m^+(t)\Psi_m(t,0)\Psi_m(0,0)d_m(0)\rangle.
\end{gather}
As is well known, the number of d electron is conserved
($[d^+d,H_{deep}]$), it is not a truly dynamic quantity, so one can
reformulate the problem in terms of time dependent core-hole potential.
After bosonization, the core level electron couples solely to the charge
sector as
\begin{equation}
H_{deep}=H+E_0\sum_md_m^+d_m+V\sqrt{\frac N\pi} \sum_md_m^+d_m\partial_x\Phi_c(0).
\end{equation}
From now on we focus on the excitation of a single $d_m$ electron, and forget about the 
presence of the other deep level particles. They can only renormalize the threshold frequency but
not the exponent what we are interested in\cite{affleck}.
Following the strategy of Schotte and Schotte\cite{schotte}, we introduce two Hamiltonians,
one when there is no hole ($H_I=H$) and another, when conduction electrons
feel a scattering potential
($H_F=H+E_0+V\sum_{m=1}^N\Psi^+_m(0)\Psi_m(0)$).
Similarly to the case of a normal Fermi liquid, there exist a unitary
transformation (the boundary condition changing operator\cite{affleck}) relating the two Hamiltonians as
\begin{gather}
H_F=U^+H_IU,\\
U=\exp\left(i\frac{2\delta}{\sqrt \pi}\Phi_c(0)\right),
\end{gather}
where
\begin{equation}
\delta=\frac{\sqrt NV}{2 v}\frac{U_0}{U_0+(N-1)U}.
\end{equation}
With the use of these,
the d electron Green's function can be calculated from
\begin{equation}
D(t)=-\langle d_m(t)d_m^+\rangle=\langle U^+(t)U\rangle.
\end{equation}
Its time dependence can easily be determined from the asymptotics of the correlator of the charge sector
using Eqs. \ref{green1} and \ref{green2} as
\begin{equation}
D(t)\sim t^{-\left(\delta/\pi\right)^2}.
\end{equation}
As is seen, the time decay implies non-Fermi liquid behaviour,
characteristic to the X-ray problem\cite{affleck}. The Fourier transform of the deep
level electron can be obtained,
from which the d-electron density of states follows as
\begin{equation}
-\textmd{Im}D(\omega)\sim
\frac{\sin(\pi(1-(\delta/\pi)^2)}{|\omega-E_0|^{1-(\delta/\pi)^2}}.
\end{equation}
The original Dirac delta density of states situated at $E_0$ transforms into a 
power-law divergence with a finite tail.
Compared to that of a normal metal without Wolff impurity, the exponent
 $\delta$ decreases for large number of channels, while in a normal metal
 the exponent grows with $\sqrt N$. Also at fixed $N$, repulsive ($U>0$)
 interaction suppresses $\delta$, while in the attractive case the exponent
 increases rapidly. It is worth mentioning, that since only the charge
 channel is involved in the X-ray Hamiltonian, the singularity present in
 the Green's function (Eq. \ref{green}) is absent here for repulsive
 interactions.
The two-particle Green's function, $S(t)$ can be evaluated similarly to
 $D(t)$, namely
\begin{equation}
S(t)=\langle U(t)\Psi_m(t,0)\Psi_m^+(0,0)U^+\rangle.
\end{equation}
Using the asymptotic formula for the correlator of the $\Phi$ field, it reads as
\begin{equation}
S(t)\sim t^{-((1-\delta/\pi)^2+N-1)/N}.
\end{equation}
Since only the charge degree of freedom is involved in the conduction-deep level electron interaction, the 
scattering phase shift comes with an additional factor of 1/N, overwhelmed by the spin degrees of freedom 
with zero phase shift.

Similar phenomenon occurs in the X-ray edge absorption, when we extend the local correlation
of our model to the whole lattice, namely in the SU(2) Hubbard model.
The X-ray exponent was found to decrease for repulsive
interaction\cite{affleck}, and increase for attractive one. 
We believe, that the qualitative dependence of local quantities on electron
correlation is 
well described by the single impurity Wolff model.

\section{Conclusions}

We have studied the low energy properties of the SU(N) Wolff impurity model via Abelian bosonization and 
the path integral treatment of the generating functional. The system decouples into N-1 identical spin 
sectors and one single charge sector, involving only chiral bosonic fields. The single particle Green's 
function possesses Fermi liquid type asymptotics, but its weight can strongly be modified by the local 
Hubbard interaction. The spin and charge sectors become inequivalent, signaling a transition to the strong 
coupling phase at $-U_0/(N-1)$ for attraction and at $U_0$ for repulsion. For small interaction, the system 
exhibits local Fermi liquid behaviour\cite{hewson} in all evaluated quantities.

When interaction with deep level electrons is considered, the X-ray response of the model can be explored.
It can be formulated similarly to the original work of Schotte and Schotte\cite{schotte,nersesyan}, 
exploiting the fact, that the number of deep level electrons is a conserved quantity. The deep level 
electron Green's function can be obtained using the boundary condition changing operators\cite{affleck}, 
reflecting the influence of both the local Hubbard interaction and the presence of N species of electrons.
The X-ray exponent is decreased
(increased) for repulsive (attractive) Hubbard interaction.

\begin{acknowledgments}
We would like to thank G. Zar\'and for valuable discussion on bosonization and on related subjects, and A. 
Virosztek for useful conversations. 
This work was supported by the 
Magyary Zolt\'an postdoctoral
program of Magyary Zolt\'an Foundation for Higher Education (MZFK).
\end{acknowledgments}
\bibliographystyle{apsrev}
\bibliography{wboson}
\end{document}